\documentclass[runningheads]{llncs}

\usepackage{graphicx}
\usepackage{amssymb,amsmath,color}
\usepackage{algorithm}
\usepackage{algorithmicx}
\usepackage[noend]{algpseudocode}
\usepackage{enumitem}

\usepackage{url}

\newcommand{\AStrats}{\mathcal{F}_{\mut(\syscorrect)}}
\newcommand{\DStrats}{\mathcal{G}_{\auditset}}

\newcommand{\soft}{{\cal S}}
\newcommand{\sysbrief}{{\cal P}}

\newcommand{\sysparam}[1]{\sysbrief[{#1}]}
\newcommand{\sys}{\sysparam{\soft}}
\newcommand{\softprime}{{\cal S}'}
\newcommand{\sysprime}{\sysparam{\softprime}}

\newcommand{\syscorrect}{\sys}

\newcommand{\result}{\mathit{result}}
\newcommand{\executions}{\mathit{exec}}

\newcommand{\audit}{\mathit{a}}
\newcommand{\auditset}{\mathcal{A}}
\newcommand{\mut}{m}

\newcommand{\In}{\mathit{In}}
\newcommand{\para}[1]{\medskip\noindent\textbf{#1}}

\newcommand{\dott}{\;.\;}

\newcommand{\Results}{\mathit{AccResults}}
\newcommand{\SR}{\mathit{SR}}
\newcommand{\prob}{\mathit{Pr}}

\newcommand{\votes}{v}

\newcommand{\rvar}[1]{\mathsf{#1}}

\newcommand{\outcome}{\omega}
\newcommand{\outset}{\mathcal{W}}

\newcommand{\siSAS}{1}
\newcommand{\siSASWJ}{2}
\newcommand{\siWJ}{3}
\newcommand{\siProbAud}{4}

\newcommand{\comment}[1]{}
\definecolor{tucgreen}{RGB}{0,140,79}
\newcommand{\new}[1]{{\color{tucgreen}#1}}
\renewcommand{\new}[1]{#1}
\renewcommand{\note}[1]{}

\title{A Declaration of Software Independence}

\author{
     Wojciech Jamroga\inst{1}, Peter Y.A.~Ryan\inst{1}, Steve Schneider\inst{2}, \\ Carsten Sch\"{u}rmann\inst{3},
    Philip B.~Stark\inst{4} }

\authorrunning{W. Jamroga et al.}

\institute{
University of Luxembourg 
\and
University of Surrey
\and
IT University of Copenhagen
\and
University of California, Berkeley
\\
Authors listed alphabetically}
\date{}

\begin{document}

\maketitle

\begin{abstract}
A voting system should not merely report the outcome: it should also provide sufficient evidence to convince reasonable observers that the reported outcome is correct.
Many deployed systems, notably paperless DRE machines still
in use in US elections, fail certainly the second, and quite possibly the first
of these requirements. 
Rivest and Wack proposed the principle of \emph{software independence} (SI) as a
guiding principle and requirement for voting systems.
In essence, a voting system is SI if its reliance on software is ``tamper-evident'', that is, if there is a way to detect that material changes were made to the software without inspecting that software. 
This important notion has so far been formulated only
informally.

Here, we provide more formal mathematical definitions of SI.
This exposes some subtleties and gaps in the original definition, among them:
what elements of a system must be trusted for an election or system to be SI,
how to formalize ``detection'' of a change to an election outcome, the fact that SI is with respect to a set of detection
mechanisms (which must be legal and practical),
the need to limit false alarms,
and how SI applies when the social choice function is not deterministic.

\end{abstract}

\section{Introduction}\label{sec:intro}

Using digital technologies in elections opens up possibilities of enriching democratic processes, but it also brings  a raft of new and often poorly understood threats to election
accuracy, security, integrity, and trust.
This is particularly clear with the so-called \emph{DRE}, Direct-Recording Electronic
voting machines, deployed widely in the U.S.~after the Help America Vote Act (HAVA) of 2002, which passed in the aftermath of the controversial 2000 presidential election.
The original DREs recorded, reported, and tallied cast votes using just software,
with no paper record.
Thus, an error in or change to that software could
alter the outcome without leaving a trace.

It might be argued that the software could be analysed and proven to always deliver a correct
result given the input votes.
In practice, such analysis and testing is immensely difficult and prohibitively expensive.
Moreover, access to the code is often restricted due to commercial or legal constraints.
And even if the software could be analysed completely
and proven correct, there is still the challenge of guaranteeing that the software actually
running on all the machines throughout the voting period is the ``correct'', verified version.

Consequently, for paperless DRE machines, BMDs, and existing Internet voting systems, voters, election officials et al.\ are required to place total blind
confidence in the correctness of the code running on the devices.

Such concerns prompted calls to add a
Voter-Verifiable Paper Audit Trial (VVPAT) to DREs, essentially a printer attached to the DRE that
prints the voter's choice, in sight of the voter.
In principle, each voter can check whether the paper accurately recorded her preferences, and correct the record if not.\footnote{There is considerable evidence that
voters rarely check machine-generated printout and are unlikely to notice that votes were altered. See, e.g.,
\cite{everett07,deMilloEtal18,bernhardEtal20,haynesHood21}.
}

An alternative response---piloted but not yet widely adopted for political elections---is cryptographic end-to-end verifiable voting (E2E-V), which provides voters a means to verify that their vote reaches the tally unaltered and is correctly included in the tally. 
An accessible introduction to such systems can be found at \cite{bernhardEtal17}, and a more extensive description at \cite{hao2017}

To capture the essential goal of being able to detect whether faulty software altered the outcome while remaining agnostic with respect to the technology employed to achieve that goal 
(e.g., a paper record or cryptographic methods), \cite{rivestWack06,rivest2008notion} proposed the principle
of \emph{software independence},
which seeks to exclude systems for which the trust in the correctness of the outcome requires trusting the software.
The original definition is given as follows:
\begin{quote}
A voting system is \emph{software-independent} if
an (undetected) change or error in its software cannot
cause an undetectable change or error in an election outcome.
\end{quote}

\cite{rivestWack06,rivest2008notion} also define a stronger requirement, a system that does not require trusting software, and that is resilient to software-caused errors:

\begin{quote}
A voting system is \emph{strongly software-independent} if it is software
independent
\emph{and moreover, a
detected change or error in an election outcome (due to change or
error in the software) can be corrected without re-running the
election}.
\end{quote}

Version~2.0 of the U.S.~Voluntary Voting System Guidelines \cite{vvsg2}, 
adopted 10~February 2021,
incorporates the principle of Software Independence:
\begin{quote}
   9.1 - An error or fault in the voting system software or hardware
cannot cause an undetectable change in election results.
\end{quote}

The principle seems very natural and compelling.
It clearly rules out paperless DRE machines and---subject to
certain assumptions about voter eligibility and the curation of paper ballots---it clearly admits systems based on hand-marked paper ballots supporting manual recounts, risk-limiting audits
\cite{stark2008conservative}, and other forms of audits.
However, as soon
as we start to consider applying it to other systems, such as end-to-end cryptographically verifiable systems, things are less clear. 
In particular,
many of the terms used in the definition require careful interpretation:

\begin{itemize}
\item What exactly do we mean by \emph{the system}?
Does it include pollworkers? Auditors? Where do we draw the boundaries?

\item What exactly is the \emph{software}?
Does it include software involved in determining voter eligibility? Auditing software?

\item What exactly does it mean to \emph{detect} an error? 
Is it enough simply to flag a problem, or must evidence be provided that there really is a problem? 
What kind of evidence?
To whom is the evidence available \cite{appelEtal20}? 
What rules out systems that always cry ``foul'', even when the election outcome is correct?

\item What do we mean by \emph{outcome}, in particular, where the social choice function is non-deterministic?
\end{itemize}

All of this motivates a more formal statement of the principle, which is the aim of this paper.
This reveals a number of subtleties, notably that the original definition, read literally, does not exclude systems that reject every
declared outcome: there is no penalty for false alarms. We argue that while software independence is a necessary property for a system to be able to deliver a verifiable outcome,
it is not sufficient. 
We also stress the distinction between a \emph{system} being \emph{verifiable} and an \emph{election} being \emph{verified}.

We do not here address vote anonymity, receipt-freeness, coercion resistance, and related concerns.
We focus just on the issues of detecting and correcting
wrong outcomes while controlling false alarms.
In practice, of course, great care needs to be taken in designing a system to
provide sufficient transparency and generate sufficient evidence without violating privacy requirements.

We should also remark that, while software independence means that we should not have to place blind faith in the correct behaviour of the code, this does not imply that we should do away with all verification and testing of code. The latter is still important to help ensure the smooth running of any election run using the system, but the assurance of the outcome should not depend on the rigor etc. of such measures.

SI is a desirable property,
but the use of an SI system does not by itself
give the public adequate reason to trust election outcomes. 
The fact that it is possible to detect malfunctions of the software does not mean that checks will be performed nor that appropriate action will be taken if problems are detected.
And errors or corruption may occur outside the software, e.g. breaches of chain of custody, faulty procedures, incorrect electoral roles, etc.

The notion of software independence is related to notions of end-to-end verifiability (E2E-V);  we discuss the relationship in Section~\ref{sec:e2ev}.

\section{Formalizing Software Independence}\label{sec:setup}

In this section we set the ground for a definition that seeks to capture more formally the spirit of the original natural-language definition.
We believe it is faithful to the spirit of the original, but as we shall see, the definition reveals some subtleties, and motivates the game-theoretic definition of the notion of \emph{evidence-based elections}
\cite{starkWagner12,appelStark20}, presented below.\footnote{
   The idea of evidence-based elections is that election officials should not only find the correct winner(s),
   but should also produce convincing public evidence that they found the correct winner(s)---or else admit
   that they cannot.
}

\subsection{Software Independence... of What?}

To merit public confidence, a voting system should generate evidence that can be used to check whether it behaved correctly; typically, that involves a tamper-evident record of voters'
expressed preferences, to which the  social choice function can be applied to check the reported result.
That record might be in the form of a well curated paper audit trail, or, as in many E2E-V  systems, data (some of which is encrypted) posted to a public bulletin board (ledger). 
Furthermore, the system should provide for various checks to be performed on this evidence by the stakeholders: voters, observers, candidates etc. 
Such checks might be performed before the election starts (e.g. verifying that a transparent ballot box is initially empty), during (e.g. Benaloh challenges), or after (e.g. risk limiting-audits, risk-limiting tallies, verification of zero-knowledge proofs, digital signatures etc.).  
We refer to such checks generically as ``audits''.

We consider \emph{software independence as a property of a voting system $\sysbrief$ with respect to a set of audits $\auditset$}. 
The voting system $\sysbrief$ represents all the components and aspects relevant for how the election is run, starting with the voting protocol, including its implementation (software) and deployment (hardware, physical infrastructure), specification of the environment, assumptions about human users, threat models, etc.
The set of audits $\auditset$ captures the notion of ``detectability'' by providing an abstract representation of the methods available for detecting something is amiss.

We emphasize that it only makes sense to talk about software independence with a particular view of detection methods.
For example, a voting system might be SI if a very powerful (and expensive) kind of instrument or audit can be used, but not if the requisite tools and methods are unaffordable, too time-consuming, or not mandated in law or regulation. 
On the other hand, another voting system might not be SI with respect to any known audit method, yet may become SI if a new forensic method is invented.\footnote{E.g., think of what happened to criminal forensics when DNA tests were introduced.
}
We elaborate on both aspects of this characterisation below.

\subsection{Voting System and Its Software}\label{sec:voting-system}

Let $\sysbrief$ be a specification of how the voting protocol should work.
\new{This refers to the overall election system, including hardware, software, procedural, and human components.}
More precisely, $\sysbrief$ denotes the system running ``correct'' software, i.e., software that correctly computes the chosen social choice function over the voted preferences of eligible voters.
The software, denoted $\soft$, is considered a part of the system. However, in an actual execution of the system, $\soft$ \new{may be} under the control of the adversary.
Thus, $\soft$ denotes a part of the system on whose correct behaviour we do not want to rely for evidence that the result is correct.
In practise, that might comprise more than software. 
The spirit of the original definition corresponds to taking $\soft$ to be the
software that records and interprets votes, applies the social choice function to them, and reports an outcome.  
It does not include software that may form part of the
surrounding system, such as software involved in giving each voter the correct ballot, software used to verify voter eligibility (e.g. voter registration systems and
electronic pollbooks), or software involved in auditing the results.
Nor does it include the behavior of voters, pollworkers, or election officials.

When we want to make the software $\soft$ explicit in the voting system $\sysbrief$, then we write $\sys$.
Note that it is straightforward to generalise our approach to quantify over other parts of the system, e.g., hardware, people, procedures, etc.

The relevant aspects of system $\sys$ are characterized, on an abstract level, by the following sets and functions:
\begin{itemize}
\item $\mut(\sysbrief)$: a function that returns all the relevant mutations $\sysprime$ of the voting system $\sysbrief$. We consider $\sysprime$ as a relevant mutation of $\sys$ if $\sysprime$ can be obtained from $\sys$ through changing only the software of $\sys$, i.e., $\soft$.  
    Hardware and processes and protocols must be the same for $\sys$ and $\sysprime$.
    The software that can be changed is restricted to the software involved in collecting voter selections (votes), applying
    the social choice function to the votes,
    and reporting the results.

\item $\In$: the set of possible input sequences. Typically, an input sequence will comprise all the votes expressed\footnote{By \emph{expressed}, we mean
    what the voter did: the marks the voters make on the paper
    or the cell they press on a touchscreen.
    Of course, a confusing user interface---including poor ballot layout---can cause voters' expressed preferences to differ from their intended preferences.
    See, e.g. \cite{appelEtal20}.
    } 
by the voters. 
Depending on the level of granularity in our modelling, it may also include  other election-related activity, such as voter registration steps, eligibility verification, coercion attempts, generation of cryptographic keys, where and how each vote was cast, etc.
    It may also include the full expressed preferences of all the voters.
    In general, $\votes\in\In$ contains much more information than is needed to determine who won.

\item $\Omega$: the set of possible election outcomes. Typically, an outcome is either the tally, or the winner(s) of the election. 
Depending on the level of granularity, it may also include any other publicly available output of the voting system, such as the content of the web bulletin board. %
    \new{We assume that $\Omega$ is finite.

    For example, in a plurality contest with two contestants, A and B, the possible outcomes in $\Omega$ might be ``A wins,'' ``B wins,'' and ``A and B are tied.'' If the social choice function breaks ties, then there would be only two possible outcomes: ``A wins'' and ``B wins.''}

\item $\executions(\sys,\votes)$: a function that returns the set of all the possible executions (runs) of system $\sys$ on the sequence of inputs $\votes\in\In$.
     Any particular election system with a particular input sequence might have a number of possible executions arising from the different choices that can be made at various points of the voting protocol. 
     For example if a voter is required to provide inputs
     other than just selections (e.g., to decide whether
     to challenge an encryption, as allowed in some E2E-V protocols), then different possible executions can arise.
    In practice, there will usually be just one possible execution given $(\sys,\votes)$, but there may be boundary conditions (e.g. tie-breaking, or randomness in transferring votes) where more than one result is possible.

    Naturally, $\executions(\sysprime,\votes)$ is the set of all possible executions of the mutated system $\sysprime$ on the input sequence $\votes$.

\item $\result(E)$: the outcome of the election for execution $E$.
      \new{We lift the function to sets of executions $X$ by fixing $\result(X) = \{\result(E) \mid E\in X\}$.}
      In the case of the correct system $\syscorrect$, we would expect any outcome in $result(\executions(\syscorrect,\votes))$ to be a valid result of the election.
\end{itemize}

\new{Note that the composition $\result(\executions(\sys,\votes))$ can be seen as a generalisation of a \emph{social choice function}.}

\subsection{Available Audits}\label{sec:audit}

\new{In the process of running the election, including recording, tallying, and broadcasting the election results, the overall voting system $\sys$ generates evidence that can be used to audit the election.}
The auditing of an election may overlap with, or be completely separate from the voting procedure.
The evidence is provided as input to a decision-making process, represented by a function $\audit$, which then provides a judgement. 
The software in $\audit$ is assumed to be trustworthy.  Such an assumption of trustworthiness needs of course to be justified, and this
will usually be by arguing that, if its inputs and intended function are public, anyone who wishes to check the correctness of its outputs could write it again from scratch, or a reputable authority such as the Electronic Frontier Foundation (EFF) could provide a reference implementation.

\new{
Evidence produced in the election might include voter registration databases, poll books, physical ballots, encrypted choices, cryptographic receipts, public bulletin boards, zero-knowledge proofs (ZKPs), security videos, the condition of physical seals on ballot boxes, chain-of-custody logs, logs from telephone complaint lines or websites that record ``anomalies'' voters witnessed during the election, etc.
The evidence might not include the ``plaintext'' voter preferences and generally will not include
a voter's actual interaction with a DRE or BMD.

Some evidence generated during the election will be unreliable or unavailable.
For instance, paper ballots do not provide reliable evidence of the outcome if they might have been tampered with, replaced, augmented, or lost; or if voter eligibility checks were not sufficiently accurate.
In some E2E-V systems, plaintext votes are not available to check the correctness of the outcome; a system designed to allow voters to check that their intent was recorded correctly (e.g., using a VVPAT or through a Benaloh challenge) does not provide public evidence that voter intent was correctly recorded unless there is both evidence about the number of voters who checked, how effectively they checked, and a mechanism by which it would become known that they found errors, if they find errors.
It must be also noted that, by the time a preliminary outcome is available, evidence could be lost, altered, or counterfeited; the election officials might have reacted to some detected problems during the election; and that in turn might generate new possibilities for things to go wrong.
}

Formally, the capability of the voting authorities (possibly together with independent auditors, public observers, or with voters, e.g., in case of mechanisms for voter verification) to detect malfunctioning of the voting system is characterised by the set $\auditset = \{\audit_1, \audit_2, \dots \}$ of available audit procedures.
Let $T$ and $F$ denote the truth values of \emph{true} and \emph{false}, respectively.
Each element $\audit_i$ of $\auditset$ is a function that takes an execution $E$ of the voting system, and returns an audit judgement $\audit_i(E) \in \{T, F \}$ such that $\audit_i(E) = F$ only if there is a change or error in the election outcome.
(Below we also consider audits that have a random component, and thus have some probability of returning $T$ or $F$ for any given the voting system execution $E$.)
It is required that $\audit_i$ must be compatible with the voting system, in the sense that the judgment $\audit_i(E)$ is based entirely on the evidence available in the execution $E$ of the voting system.

\new{
For instance, $\auditset$ might include verifying poll book signatures, comparing the number of pollbook signatures to the number of votes cast in each precinct, a manual audit of results against a paper trail, checking ZKPs, checking whether digital signatures on cryptographic receipts are authentic, reviewing chain-of-custody records, inspecting equipment log files and security videos, etc.

An exemplar $\audit_i$ might specify, among its branches, ``before opening each box of ballots for central counting, check the seal on the box against a photograph of the seal taken in the polling place. 
If the seal has been disturbed, interview everyone who has had custody of the box since it was sealed, examine every ballot by hand for signs of tampering or forgery, and compare the number of ballots in the box with the number of pollbook
signatures.''
}

\section{Possibilistic Formulation of Software Independence}\label{sec:possibilistic}

The original definition of SI talks about whether a change to the result is always detectable.  This is expressed in terms of possibilities rather than probabilities.  Here we see how far we can get with expressing SI possibilistically without involving probabilities.
We will also show that it is natural to introduce probabilities into the audit process.

\subsection{Basic Formulation}

Using the notation introduced in Section~\ref{sec:setup}, the property of Software Independence with respect to a \new{particular election input $\votes$} and audit method $\audit$ can be expressed as follows:
\begin{eqnarray}
\lefteqn{SI_\siSAS(\syscorrect,\new{\votes},\audit)\iff} \label{SAS_SIexec-oneaudit} \\
&& \forall \sysprime\in\mut(\syscorrect) \dott \nonumber \\
&& \big(\forall E' \in \executions(\sysprime,\votes) \dott \exists E \in \executions(\syscorrect,\votes) \dott (\result(E) = \result(E'))\big) \nonumber \\
&& {} \lor \big(\exists E' \in \executions(\sysprime,\votes) \dott \audit(E') = F \big). \nonumber
\end{eqnarray}
The formula states that every execution of any mutation of $\sys$ gives a correct result, or else the malfunction is detectable.
More precisely,
either every execution of a mutation of $\sys$ gives a result that could have been produced by the correct software,
or there is some execution that will fail the audit.

Then, Software Independence holds with respect to a set of possible election inputs $\votes\in\In$ and allowable audit procedures $\auditset$ if there is some audit procedure $\audit\in\auditset$ such that SI holds for all possible inputs:
\begin{eqnarray}
SI_\siSAS(\syscorrect, \auditset) & \iff & 
\exists \audit \in \auditset \dott \new{\forall \votes \in \In \dott} SI_\siSAS(\syscorrect, \new{\votes}, \audit). \label{SAS_SIexec-auditset}
\end{eqnarray}

\new{Arguably, formula~(\ref{SAS_SIexec-oneaudit}) captures software independence of particular \emph{election}, given the set of votes and actual audit strategy used in the election. In contrast, formula~(\ref{SAS_SIexec-auditset}) expresses software independence of the \emph{voting system} defined by the voting infrastructure and the available audit strategies.}

\para{Remarks.}
Formulas~(\ref{SAS_SIexec-oneaudit})--(\ref{SAS_SIexec-auditset}) capture a rather weak notion of Software Independence. First, they only say that $\sysprime$ cannot undetectably add incorrect outcomes to the set of possible results of the election. However, a software mutation \emph{removing} some of the correct results may as well satisfy the conditions. We address this issue in Section~\ref{sec:resilience}.

Secondly, the formalisation is based on a weak notion of detectability. The conditions require that significant software mutations \emph{might} be detected (i.e., they are detected on some possible executions), but there is no guarantee that they can be detected for every execution that produces incorrect outcomes. 

A stronger definition of SI is obtained by replacing the right hand side of the disjunction~(\ref{SAS_SIexec-oneaudit}) as follows:
\[
\forall E' \in \executions(\sysprime,\votes) \dott \big((\exists E \in \executions(\syscorrect,\votes) \dott \result(E) = \result(E')) 
\lor (\audit(E') = F) \big).
\]

This removes the existential quantification over executions and brings $E'$ under the universal quantification. The first formalisation allows for some executions of a mutation not to be caught by an audit even if they give the wrong result.  This stronger formalisation states that any execution of a mutation that does not give the correct result should be caught by an audit.

Note also that our formalisation is focused on the potential irregularities due to software mutations. Thus, disturbances of the election outcome due to failures of hardware, dishonest voter behaviour, etc., must only be handled in $\sysprime$ if they would be caught and dealt with in the ideal system $\sys$.

\new{\para{Audit Strategies.}}
We recall that the characterisation of $\auditset$ encapsulates the audit methods that are allowable. Considerations as to what should be allowable can include what is possible, affordable (in terms of cost or time), legal (to fit with local election law, preserve the anonymity of votes, etc.), and other considerations as appropriate to the situation.  Identifying the limits of what is allowable is itself part of the consideration as to whether a system is software independent.
From a technical point of view, the definition of $\auditset$ will also need to depend on the evidence provided explicitly by the voting system.  Thus the formalisation of possible executions $E$ also constrains the audits that are possible, because $\audit$ is a function on executions: two runs giving rise to the same execution record $E$ must give the same result on audit.  For example, if the only evidence collected for audit is the set of paper ballots, then forensic analysis of the hard disks of the voting machines is outside the scope of audit.  Conversely if the audit includes the possibility of such analysis, then the evidence provided by an election run should include the relevant state of the hard disks to enable the audit function to be defined.

The sanity condition (or soundness) on an auditing mechanism $\audit$ for system $\sys$ is that any correct execution of the ideal system will verify positively:
\begin{eqnarray*}
{\textbf{sound}(\audit,\syscorrect) \iff}
 \forall \votes \in \In \dott \forall E\in\executions(\syscorrect,\votes) \dott \audit(E) = T.  
\end{eqnarray*}
Although this is not stated explicitly within the original definition of Software Independence, correct election
outcomes should not be flagged by the audit as incorrect, so we will require that every $\audit$ function in $\auditset$ be
\emph{sound}.

\subsection{Relationship to End-to-End Verifiability}
\label{sec:e2ev}

The definitions above enable us to highlight an important distinction between Software Independence and End-to-end Verifiability (E2E-V), cf.~\cite{benaloh2015end} for an introduction and~\cite{kusters2011verifiability} for a well-known formalisation.  
In particular, in a description of a system $\syscorrect$ the component ${\cal S}$ explicitly represents only the software, and the context ${\cal P}$ remains unchanged.  
This amounts to requiring that the context ${\cal P}$ is trusted in the characterisation of SI.  
However, when we consider whether the system $\syscorrect$ is end-to-end verifiable, we consider this question with respect to the entire system.  

We should note that not all formulations of E2E-V in the literature actually imply correctness of the outcome. Early formulations focused on the ability to detect the corruption of any vote between casting and input to the tally function. To achieve guarantees of correctness we also need measures to prevent ballot stuffing and ballot collisions. Taken together, these imply a bijection between the set of cast votes and the set of votes input to the tally. Here we assume a definition that does encompass these requirements, as does ~\cite{kusters2011verifiability}. Here they refer to such a strengthened notion, that does imply correctness if all verification steps give true, as \emph{global verifiability}.

To illustrate the difference, consider the following toy example,
which shows that SI does not imply 
E2E-V:
A voting system consists of a ballot box for paper ballots, a scanner, and a software component ${\cal S}$ that controls the scanner, interprets the scans, applies the social choice function to the votes, and reports the result.  
There is a trusted individual ${\cal I}$ (appointed by the Election Authority, say) who will also play a key part. 
A description of the system formulated as $\syscorrect$ would include ${\cal I}$ within the definition of ${\cal P}$.

{\bf Voting:}  To vote, voters fill out their ballot form, run it through the scanner, then drop it in the ballot box.

{\bf Tallying:} At the end of the election, ${\cal I}$ privately counts the votes from the ballot box and calculates the result $r_1$.   The electronic component ${\cal S}$ computes the result $r_2$ from the scans, and provides this result to ${\cal I}$, who privately checks whether $r_1 = r_2$.  
If so, then ${\cal I}$ reports the result.  
Otherwise an alarm is raised and an audit occurs, consisting of comparing $r_1$ and $r_2$.  if they are distinct then the audit returns the value $F$.

The system $\syscorrect$ is SI, because an undetected change in ${\cal S}$ cannot undetectably change the result, and the system meets the definition in Line~\ref{SAS_SIexec-oneaudit}.  Given a change to the software, either the resulting software still gives the same result, or the audit will return the value $F$.  Note that this relies on the honesty and correct behaviour of ${\cal I}$; this is assumed for the characterisation of SI. 

The system $\syscorrect$ is not E2E-V.  Voters are not able to check that their vote is included in the tally, and there is no check for independent observers that the tally is computed correctly.  In particular, ${\cal I}$ can simply report a different result and not raise the alarm.

One key difference is that for SI, any part of the system that is not the software is presumed to be acting as it should.  
Hence, the question is whether a change to ${\cal S}$ can change the result when ${\cal P}$ behaves correctly.

On the other hand for E2E-V we also consider that ${\cal P}$ can behave dishonestly.  
So $\syscorrect$ is not E2E-V: it is possible for the wrong result to be reported without any verification checks showing incorrect behaviour.

A further distinction is that SI makes no mention of who does the ``detecting,'' whereas E2E-V is quite explicit: each voter can perform the individual check and anyone can perform the universal check. 
The example above illustrates this point, too.

\subsubsection*{E2E-V $\Rightarrow$ SI:}

Conversely, we can reason informally that E2E-V implies SI, via a contrapositive argument as follows.  If a system with verification mechanisms is not SI, then by Definition~\ref{SAS_SIexec-auditset} for some input $v$ there is a change to the software ${\cal S}'$ that can result in an execution $E'$ with an incorrect result $result(E')$ that passes every audit $audit \in \auditset$, i.e. it produces an undetectable change to the result. 
But if the incorrectness of the result is undetectable, then the verification mechanisms cannot detect this, and hence will verify an incorrect result. 
But this means the system is not E2E-V, since E2E-V requires that if 
all potential verification steps pass\footnote{I.e., every voter checks what individual voters can check (individual verifiability), someone checks the aggregation of votes (universal verifiability), and someone checks that every vote has come from a different eligible voter (eligibility verifiability).
}
then the result is correct. Note that here we are assuming a strong notion of verifiability, such as global verifiability.

Observe that both audits and verifications can raise an alarm even when the result is correct.  
We are not concerned with this case in this section, but rather the converse case where the audits and verifications do not raise the alarm even though the result is incorrect.

\subsection{SI with Adaptive Audits}

The formalization of SI by formulas (\ref{SAS_SIexec-oneaudit})--(\ref{SAS_SIexec-auditset}) assumes that there exists a single audit strategy in $\auditset$ that can detect malfunction and/or tampering with the voting software.
Another option is to swap the quantifiers, and assume that different audit procedures may be applicable on different runs of the voting system (e.g., against different kinds of threats).
Now, SI with respect to a set of available audits becomes:
\begin{eqnarray}
\new{\lefteqn{SI_\siSASWJ(\syscorrect,\auditset)\iff \forall \votes \in \In \dott SI_\siSASWJ(\syscorrect,\votes,\auditset);} } \nonumber \\ \lefteqn{SI_\siSASWJ(\syscorrect,\new{\votes},\auditset)\iff} \label{SAS_SIexec2-election} 
\\
&& \forall \sysprime\in\mut(\syscorrect) \dott \nonumber \\
&& \big(\forall E' \in \executions(\sysprime,\votes) \dott \exists E \in \executions(\syscorrect,\votes) \dott (\result(E) = \result(E'))\big) \nonumber \\
&& {} \lor \big(\exists E' \in \executions(\sysprime,\votes) \dott \exists \audit \in \auditset \dott \audit(E') = F\big). \nonumber
\end{eqnarray}
That is, either every execution of any mutation of $\sys$ gives a result that could have been produced by the correct software,
or there is some execution that will fail at least one audit procedure in the available audit set.
\new{Again, formula~(\ref{SAS_SIexec2-election}) captures software independence of an election, and~(\ref{SAS_SIexec-auditset}) expresses SI of the voting system.
Note that these notions of detection are still somewhat weak in that they do not ensure that anyone can tell \emph{which} $a \in \auditset$ suffices for any particular execution $E$.}

\subsection{A Refinement}\label{sec:refinement}

Audit procedures are often nondeterministic by design (e.g., audits that inspect a random sample of ballots, including risk-limiting audits).
In our definition of SI, it can be beneficial to separate the randomness of the audit from randomness in the rest of the system.
This view can be incorporated by 
treating audit procedures as functions on system executions $E$ that return a probability distribution on $\{T, F\}$.

For example, for statistical audit of the paper trail, different audit runs result from inspecting different random samples of ballots, each of which has some probability;
for some runs, the audit might return $T$ and for others $F$. 

The soundness sanity condition on the auditing mechanism $\audit$ stays as before.

Having separated the audit non-determinism from the system non-determinism, we can now redefine ``undetectable change'' to apply to those \emph{system runs} for which the probability that the audit returns $F$ is zero.
Let  $\prob$ denote probability computed with respect to the audit, treating .
Now, software independence of system $\syscorrect$ with respect to the audit set $\auditset$ becomes:
\begin{eqnarray}
\lefteqn{SI_\siWJ(\syscorrect,\auditset) \ \,\iff \ \exists \audit \in \auditset \dott \new{\forall \votes \in \In \dott} SI_\siWJ(\syscorrect, \new{\votes}, \audit);} \label{WJ_SI-auditset} 
\\
\lefteqn{SI_\siWJ(\syscorrect, \new{\votes}, \audit) \iff} \label{WJ_SI-audit} 
\\
&& \forall \sysprime\in\mut(\syscorrect) \dott \forall E' \in \executions(\sysprime,\votes) \dott \nonumber \\
&& (\exists E \in \executions(\sys,\votes)\dott  \result(E) = \result(E')) \lor \nonumber \\
&& \prob (\audit(E') = F) > 0.\qquad\qquad\qquad\qquad\qquad\qquad\mbox{} \nonumber
\end{eqnarray}

The definition can be equivalently phrased as follows. Let
\begin{eqnarray*}
&& \Results(\sysprime,\audit,\votes) = \left \{\outcome\ \mid\ \exists E'\in\executions(\sysprime,\votes) \dott \right .\\
&& \qquad \left . \left ( \outcome=\result(E') \land 
\prob(\audit(E') = T) = 1 \right ) \right \}
\end{eqnarray*}
be the set of \emph{surely accepted results} for $\sysprime$ on $\votes$. 
That is, these are the possible outcomes of running $\sysprime$ on input $\votes$ for which the audit has zero probability of reporting that the outcome is wrong.
Note that, for the ideal system $\syscorrect$, if the audit meets the soundness condition
this is just the set of possible (correct) outcomes, i.e.,
$\Results(\sys,\audit,\votes) = \{\result(E)\ \mid\ E\in\executions(\sys,\votes) \}$.
\new{Since in that case the set does not depend on the audit strategy, we will often write $\Results(\sys,\votes)$ instead of $\Results(\sys,\audit,\votes)$.
Then, formula~(\ref{WJ_SI-audit}) can be rephrased as:
\begin{eqnarray*}
\lefteqn{SI_\siWJ(\syscorrect,\votes,\audit)\iff } \\
&& \forall \sysprime\in\mut(\syscorrect) \dott \Results(\sysprime,\audit,\votes) \subseteq \Results(\syscorrect,\votes). \nonumber
\end{eqnarray*}
}

\subsection{Software Resilience}\label{sec:resilience}

The above definition says that every execution of $\sysprime$ either simulates a legitimate execution of $\syscorrect$ or has a strictly positive chance of being ``detected'' by the audit.
This kind of property is arguably closest to the spirit of the proposal by Rivest and Wack.
Also, it corresponds to the intuition that, usually, the only evidence that one has to determine a property of an election system comes from the actual run of the system during the actual election.
However, as a system property, it is rather weak.
Ideally, one would also like to guarantee the ``vice versa'' condition, saying that every outcome of the ideal software can be produced by the mutation $\sysprime$.
That is, $\sysprime$ not only does not introduce any illegal winners, but also does not remove any legally possible ones.
Then, every mutation $\sysprime$ must produce exactly the same set of acceptable election outcomes as the ideal system $\sys$.
We call the new property \emph{software resilience (SR)}, and define it formally as follows:
\begin{eqnarray*}
\new{\lefteqn{\SR(\syscorrect,\auditset)\iff \exists \audit \in \auditset \dott \forall \votes \in \In \dott \SR(\syscorrect, \votes, \audit);}} \\
\lefteqn{\SR(\syscorrect, \votes, \audit) \iff} \\
&& \forall \sysprime\in\mut(\syscorrect) \dott \Results(\sysprime,\audit,\votes) = \Results(\syscorrect,\votes). \end{eqnarray*}
In other words, $\SR(\syscorrect, \votes, \audit)$ requires that every mutation $\sysprime$ is trace-equivalent to $\syscorrect$ with respect to the surely accepted election outcomes that they can produce.

In practice of course, what the electorate needs is a way to determine, as the end of a given election, whether the reported outcome was not only one of the possible correct outcomes, but also fair in some sense. Where the outcome is uniquely defined this is fine: it is enough that we can determine that it was correct. Where the outcome is not uniquely defined, for example in the event of a tie in a simple plurality vote resolved by the system's software (rather than, for instance, by a public coin toss), this is more delicate: we would like to be able to establish that no possible outcomes were excluded by that particular software running at the time. If the tie is resolved by the software, there is no way to establish one the basis of observation of a single run.

In order to resolve such situations it seems necessary to externalise the mechanism that makes the choice amongst possible outcomes, for example based on a publicly observable coin toss or equivalent. How to provide a truly random source that cannot be predicted or influenced by any way is a topic in its own right, outside the scope of this paper.

Another approach is to regard the outcome as the raw tally, and the resolution of any ties etc. to be outside the scope of the definition. 
However, the outcome can be correct even when the tally is not---indeed, this is why risk-limiting audits can be efficient.
Machine tallies of hand-marked paper ballots are rarely if ever perfectly accurate.

Moreover, non-determinism may be buried in the tabulation algorithm itself, and so not neatly separable.
This is for instance the case in the STV variant used in New South Wales, Australia, as well as the D'Hondt method of allocating seats in the parliament in many European countries.

\subsection{Thought Experiment}

A simple voting system with rather a weak audit highlights some aspects of the definitions.

Consider a voting system $\sysbrief_{\mathit weak}$ defined as follows:

\subsubsection*{Voting}
\begin{enumerate}
\item Votes are cast on paper (filling in a bubble by hand), scanned, and then
  deposited into a ballot box.
  The scans are linked to the corresponding paper ballots in a way that allows the scan corresponding to a particular ballot to be retrieved, and vice versa.
\item All of the scans are then published, and the result declared.
\end{enumerate}
Here the software $\soft$ controls the scanning, tabulation, and reporting.
We assume that there is good physical security of the ballots, and that the total number of ballots is known.
\subsubsection*{Audit}
\begin{enumerate}
\item Auditors check whether the number of scans matches the number of ballots. If not, the audit returns $F$.
\item Auditors inspect every scan and tabulate the resulting interpretation of the votes to obtain an electoral outcome.
If that outcome differs from the reported outcome, the audit returns $F$.
\item A paper ballot is selected at random. Its corresponding scan is retrieved and checked to
 see whether the human interpretation of that scan matches the human interpretation of the ballot. 
 If not, the audit returns $F$. 
\end{enumerate}

According to the Rivest/Wack definition of SI this system is SI, because any change in the result (caused by a change in the software) can be in principle detected. Thus, it meets the formal characterisation in Line~\ref{SAS_SIexec-oneaudit}.
However, this audit may have a low probability of
detecting an attack that alters or substitutes scans.  
If the fraction of the altered scans is $\delta$, then $\delta$ is also the chance of detecting the attack.
(Moreover, this audit may produce false alarms: the reported outcome could be correct even if some scans were altered.)

\subsection{Software Independence for Probabilistic Audits}\label{sec:probaudits}

The thought experiment illustrates that audits can be (and usually are) probabilistic.
Although the Rivest/Wack definition of software independence is expressed in possibilistic terms, a comment (almost in passing) in \cite{rivest2008notion} indicates that in practice there should be a high probability of detecting software misbehaviour:
\begin{quote}
The detection of any software misbehavior does not need to be perfect; it only
needs to happen with sufficiently high probability, in an assumed ideal environment
with alert voters, pollworkers, etc.
\end{quote}
This is a rather stronger requirement, and introduces probability into the characterisation. 
Where should this probability be introduced?  %

The idea should be that whatever mutation of $\sysbrief$ is considered, and for any execution of that mutation, if the result has been changed then this should be detectable with high probability. 
The `detectable' element of this definition is the responsibility of the audit function.

Then we can adjust the definition of Software Independence \new{of Section~\ref{sec:refinement}} to incorporate the additional requirement that when the result has been changed, the audit has a probability $p_0 > 0$ to notice that:
\begin{eqnarray}
\lefteqn{SI_{\siProbAud}(\syscorrect,\auditset,p_0)\iff
\exists \audit \in \auditset \dott \new{\forall \votes \in \In} \dott SI_{\siProbAud}(\syscorrect, \new{\votes},\audit,p_0);} \nonumber
\\
\lefteqn{SI_{\siProbAud}(\syscorrect, \new{\votes}, \audit,p_0)\iff} 
\nonumber
\\
&& \forall \sysprime\in\mut(\syscorrect) \dott \forall E' \in \executions(\sysprime,\votes) \dott \nonumber \\
&& \qquad (\exists E \in \executions(\sys,\votes) \dott \result(E) = \result(E'))\qquad\qquad\qquad\qquad\mbox{} \nonumber \\
&& \qquad {} \lor \prob (\audit(E') = F) \ge p_0. \nonumber
\end{eqnarray}
This is clearly stronger than the previous definition in \new{Equations~(\ref{WJ_SI-auditset})--(\ref{WJ_SI-audit})}.

\section{Probabilistic/Game-Theoretic Definition}

\new{
In the previous section, we proposed a possibilistic definition of software independence. 
It was based on the assumption that we can quantify over possibilities (possible mutations of the software, executions of the system, etc.) but cannot formulate constraints with respect to quantitative measures over the possibilities (e.g., probability of executions or computational complexity of a mutation strategy).
The first step towards a more quantitative approach was discussed in Section~\ref{sec:probaudits} where we considered audits with a random component.
Here, we present a full-blown quantitative definition of SI.
We assume the following:
\begin{enumerate}
\item The execution of $\sys$ on an input $\votes$ defines a probability distribution over all the possible runs in $\executions(\sys,\votes)$;
\item The execution of audit method $\audit$ given a system execution $E$ defines a probability distribution on $\{T, F\}$;
\item The choice of a software mutation belongs to a potentially malicious ``attacker,'' whereas the auditing method is selected by the ``defender.'' 
The input sequence $\votes\in\In$ is chosen by Nature;

\item The defender must select the audit without knowing the mutation the attacker selected. (However, the audit procedure can be adaptive.) The attacker knows the defender's audit strategy in advance, but not any random elements involved in that strategy.
E.g., the attacker might know that the auditor will examine a random sample of ballots, but does not know which particular ballots will be examined.
\end{enumerate}
}

\subsection{Terminology and Notation}

As before, $\prob$ denotes probability.
Moreover, we will use $\rvar{Exec(\sys,\votes)}$, $\rvar{Res(E)}$, and $\rvar{Aud(E)}$ for the random variables ranging over possible runs $E\in\executions(\sys,\votes)$, possible election outcomes $\outcome\in\result(E)$, and audit judgments in $\{T,F\}$, respectively.

\new{
\para{Election Environment.}
Given the input $\votes\in\In$ (in particular, the voters' expressed preferences), the voting system $\sys$ defines a probability distribution $\prob(\rvar{Exec(\sys,\votes)}=E)$ over the possible runs $E\in\executions(\sys,\votes)$.
Similarly, given a run $E$ of the voting system, $\prob(\rvar{Res(E)}=\outcome)$ denotes the probability that the election outcome is $\outcome\in\Omega$.
Note that the social choice function can be now represented by the probability distribution 
\[\prob(\rvar{Res(\sys,\votes)}=\outcome) = \sum_{E\in\executions(\sys,\votes)}\prob(\rvar{Exec(\sys,\votes)}=E)\cdot\prob(\rvar{Res(E)}=\outcome).\]
Deterministic social choice functions amount to randomized functions that put all their mass on a single $\outcome \in \Omega$.

For instance, in a two-candidate plurality contest with ties broken at random, the set of outcomes can be defined as $\Omega = \{a,b\}$ with $a$ standing for ``Alice wins'' and $b$ for ``Bob wins.''
If the election input $\votes\in\In$ contains more votes for Alice than for Bob, then $\prob(\rvar{Res(\sys,\votes)}=a)=1$ and $\prob(\rvar{Res(\sys,\votes)}=b)=0$.
If $\votes$ contains more votes for Bob than for Alice, then $\prob(\rvar{Res(\sys,\votes)}=a)=0$ and $\prob(\rvar{Res(\sys,\votes)}=b)=1$.
If $\votes$ has the same number of votes for Alice and Bob, then $\prob(\rvar{Res(\sys,\votes)}=a)= \prob(\rvar{Res(\sys,\votes)}=b)=\frac{1}{2}$.
}

If an election produces outcome $\outcome$ that has probability zero, that is, if $\prob(\rvar{Res(\sys,\votes)}=\outcome) = 0$, then the outcome is presumptively incorrect.\footnote{Recall that the set of outcomes is assumed to be finite.}
For a single election, if $\prob(\rvar{Res(\sys,\votes)}=\outcome) > 0$, we cannot tell whether $\sys$ assigns the \emph{correct} probability to $\outcome$: that would require replicating the execution.
Hence, we consider an outcome $\outcome$ to be \new{\emph{admissible} for $\sys$ and $\votes$} if the probability of that outcome is strictly positive, that is, if $\prob(\rvar{Res(\sys,\votes)}=\outcome) > 0$ (the outcome is expected to occur sometimes for that vote profile and that social choice function). \new{We denote the set of such outcomes by $\outset_{\sys,\votes}$.} 

\para{Attack and Defense Strategies.}
We model the interplay between \new{threats} (regardless of their cause) and mitigations as the election unfolds by means of two \emph{strategies} that play against each other: an \emph{attack strategy} and a \emph{defense strategy}.

\new{
An \emph{attack strategy} $f$ interferes with the ideal operation of the election by changing the ``software''  of the election system.
(Recall that we use
the term ``software'' abstractly, to denote those things under consideration that might behave incorrectly, which might include more than computer code, depending on context.)
Each $f$ amounts to a (possibly randomized) plan that specifies the action that the attacker will take if a given circumstance occurs.
It involves the vulnerabilities and failure modes of the overall election, and represents how outcomes and evidence might be altered by failures or adversarial attacks.
The involved software mutations are drawn from $\mut(\syscorrect)$.
}
The input $\votes$ is the set of ``true'' votes of the eligible voters.

\new{We denote the set of \emph{feasible attack strategies} by $\AStrats$.
Note that such strategies may have to satisfy some constraints.
}
For instance, it might not be computationally feasible to fake a ZKP.
Or it might not be possible to alter marks on paper ballots undetectably, to steal a ballot box and its contents undetectably,
or to corrupt a multipartisan group of auditors into faking audit results.

A \emph{defense strategy} $g$ conducts tests and countermeasures to judge whether the announced outcome of the election is correct.
Each $g$ amounts to a (possibly randomized) \new{conditional plan} that specifies the actions the defender will take in a given set of circumstances. Defense strategies consist of actions that the ``checkers'' (elections officials, auditors, public, etc.) can take before, during, and after the election to try to ensure that the outcome is correct, and to assess whether the outcome is correct, despite the fact that things might have gone wrong---that is, despite $f$. 
Clearly, they can have random elements, such as statistical audits.
Given an election run $E$, $\prob(\rvar{Aud(g,E)}=AJ)$ is the probability that the defense strategy $g$ returns audit judgment $AJ\in\{T, F\}$ on $E$.
\new{The set of possible defense strategies based on audit methods $\auditset$ is denoted by $\DStrats$} 
The set $\DStrats$ is fixed after $\AStrats$ is known, but before the apparent outcome $\outcome$ is known, and without knowledge of $f$.
That is, methods for assessing the outcome may depend on the kind of evidence the system generates, the ways the ideal evidence might be corrupted, and the execution trace $E$, including reported tallies and outcomes.
\new{The strategies in $\DStrats$ must satisfy  legal and practical constraints, as discussed
above.}

Both $f$ and $g$ are ``interactive,'' in the sense that the actions taken under a particular $g$ can depend on circumstances generated by the actions under $f$, and \emph{vice versa}, as well as on random elements.
The defense strategy is restricted to the ``audits''; the attacker has no influence on audits other than through $\soft$.

\new{
\para{Execution Semantics for Strategies.}
The choice of attack ($f$) and defense ($g$) strategies determine how probable different election runs are, which in turn affects the chance that the audit identifies incorrect outcomes. 
We model this through the probability distribution
$\prob(\rvar{Exec(\sysparam{\soft,f},\votes)}=E)$ on the set of system executions, for system software $\soft$, attack strategy $f$, and input votes $\votes$.  
For any given $g$, this induces a probability distribution on the
audit decisions $\rvar{Aud(g,E)}$. Now,
\begin{eqnarray*}
\lefteqn{\prob(\rvar{Aud(f,g,\votes)} = AJ \mid \outset) =} \\
&& \sum_{\outcome\in \outset}\ \sum_{E\in\executions(\sysparam{\soft,f},\votes)}\ \prob(\rvar{Exec(\sysparam{\soft,f},\votes)}=E) \cdot \prob(\rvar{Res(E)}=\outcome) \cdot \prob(\rvar{Aud(g,E)}=AJ) \nonumber
\end{eqnarray*}
denotes the probability that the announced outcome will be accepted (for $AJ=T$) or rejected (for $AJ=F$), given that the announced outcome is in $\outset$.
}

As in Section~\ref{sec:possibilistic}, we take $\votes$ to be fixed when defining software independence of a particular \emph{election}. 
Moreover, we are interested in $\outset=\{\outcome\}$, where $\outcome$ is the outcome that has been announced.
In defining software independence of an \emph{election system}, we quantify over the possible election inputs $\votes\in\In$, and do not condition on $\outset=\{\outcome\}$.

\subsection{Game-Theoretic Definition of SI}

We will cast software independence in terms of a game, in a manner analogous to how semantic security of cryptographic algorithms is captured, or to how estimation problems are formalized in
statistical decision theory.
\new{An election is seen as a strictly competitive game between the adversary choosing an attack strategy $f\in\AStrats$ and the checker choosing a defense strategy $g\in\DStrats$. 
The payoffs of the checker are multicriterial (and thus only partially ordered), and given by the respective probabilities of false positive and false negative output of the audit procedure. The solution concept is based on \emph{minimax}, i.e., the checker minimizes the loss assuming the worst case (most damaging) of the adversary. (Since the payoff is multicriterial, there is no minimax strategy \emph{sensu stricto}, but the analysis is worst-case.) 
Moreover, the adversary is assumed to adapt the attack strategy $f$ to the defense strategy $g$ selected by the checker. 
On the other hand, the checker must choose the defense strategy without knowing the attack strategy. }

Formally, given a defense strategy $g \in \DStrats$, \new{an election input $\votes\in\In$, and a set of admissible election outcomes $\outset_{\sys,\votes}$}, we define two kinds of costs that the checker wants to minimize:
\begin{eqnarray*}
\epsilon(g, \votes) &=& 
    \sup_{f \in \AStrats} \prob(\rvar{Aud(f,g,\votes)} = F \mid \outset_{\sys,\votes}), \\
\delta(g, \votes) &=&
    \sup_{f \in \AStrats} \prob(\rvar{Aud(f,g,\votes)} = T \mid \overline{\outset}_{\sys,\votes}) \\
    &=&    1 - \inf_{f \in \AStrats} \prob(\rvar{Aud(f,g,\votes)} =F \mid \overline{\outset}_{\sys,\votes}). \end{eqnarray*}
That is, $\epsilon$ is the largest chance that the checker rejects an admissible outcome \new{(\emph{false negative})}, and $\delta$ is the largest chance that he fails to reject an inadmissible outcome \new{(\emph{false positive})}.

\begin{definition}[$(\epsilon, \delta)$-SI]
\new{Consider an election where $\votes$ was the actual input and $g$ the used defense strategy. 
The election is \emph{$(\epsilon_0, \delta_0)$-software independent}} if $\epsilon(g, \votes) \le \epsilon_0$ and $\delta(g, \votes) \le \delta_0$, \new{i.e., the probability of false negative is bounded by $\epsilon_0$, and the probability of false positive is bounded by $\delta_0$}.

Moreover, the \emph{voting system} is $(\epsilon_0, \delta_0)$-software independent if there exists $g \in \DStrats$ such that for all $\votes\in\In$, the resulting election is $(\epsilon_0, \delta_0)$-SI.
\end{definition}

\new{
Ideally, elections should be fully reliable. This motivates the following definition.
\begin{definition}[Strict SI]
An election (respectively, voting system) is \emph{strictly software independent} if it is $(0,0)$-software independent.
\end{definition}
}

\new{Unfortunately, strict SI might be hard to achieve in realistic scenarios. In that case, we should at least require that the defense strategy is more effective than random guessing.}
Suppose that the checker tosses a biased coin (independently of all other election processes)
that has probability $p$ of landing heads, and then rejects the
announced outcome if the coin lands heads and accepts the outcome if the coin lands tails.
That rule $g_p$ attains $\epsilon(g_p, v) = p$ and $\delta(g, v) = 1-p$, so $\epsilon(g_p, v) + \delta(g_p, v) = 1$. By using the available evidence one should be able to do better.
This leads to the following definition:

\begin{definition}[\new{loose SI}]
An election (respectively, voting system) is \emph{loosely software independent} if it is $(\epsilon, \delta)$-software independent with $\epsilon + \delta < 1$.
\end{definition}

For example, consider a voting system based on hand-marked paper ballots kept secure and trustworthy, with trustworthy eligibility determinations, subject to a risk-limiting audit with risk limit $\alpha < 1$.
Such a voting system is $(0, \alpha)$-SI and loosely SI.
If there were an automatic recount instead of a risk-limiting audit,
the system would be strictly SI.

\section{Conclusions}

We have presented several formalisations of the notion of software independence. In doing so we have shown that, like many security properties, this seemingly simple and intuitive notion actually harbours many subtleties. For example we observe that it is important to exclude trivial systems that simply reject all runs of an election. The original definition clearly intended this but did not explicitly require it. Many of the terms used in the definition require precise definition. For example, ``detection'' should not just mean claiming to have observed a departure from correct behaviour but also to be able to provide evidence that such a departure did indeed occur. This is related the notion of dispute resolution: the ability of a third party to be able to determine whether alarm is genuine or false. 

We have enriched our definitions to allow for non-determinism or randomisation in the execution of the protocols, and in particular in the social choice function. Further, we have argued that purely possibilistic definition is not necessarily that useful, rather one should extend that definition to account for the probabilities of detecting erroneous behaviour.

Another insight from our formalisation is the need to precisely define when is meant by the ``system'' and the ``software''. By the latter we mean those parts of the system on whose behaviour we do not want the correctness of the outcome to depend. However, for many systems this will not include all the software of the system, for example, the auditing components and procedures may require software and we typically assume that such software is correct with respect to its specification. Such assumptions can typically be justified by arguing that  auditing algorithms can typically be rerun on independent implementations, so corruption of an instance of this software is itself detectable.

In future work we plan to apply our definitions to a representative sample of verifiable voting systems. We also plan to generalise the notion of software independence to include other components of the system: hardware, people, procedures etc. This brings us back to the question of defining the boundaries of the sub-system that we require the correctness of the outcome to be independent.

\subsection*{Acknowledgements}

Peter Y.A. Ryan would like to thank the FNR (Fond Nationale de Research Luxembourg) and the Velux Foundation for support during his sabbatical and to ITU Copenhagen for hosting him when this work was initiated. Steve Schneider is grateful to EPSRC for funding through the VOLT project EP/P031811/1.
Wojciech Jamroga acknowledges the support of the National Centre for Research and Development, Poland (NCBR), and the FNR Luxembourg under the PolLux/FNR-CORE projects VoteVerif (POL\-LUX\--IV/1/2016) and STV (POLLUX-VII/1/2019).

\bibliographystyle{alpha}
\newcommand{\etalchar}[1]{$^{#1}$}

\end{document}